\documentclass{PoS1}
\title{Charm Leptonic and semi-Leptonic decays at  BESIII}

\ShortTitle{Charm Leptonic and semi-Leptonic decays at BESIII}

\author{\speaker{Jiangchuan Chen}\thanks{Work supported in part by
the National Natural Science Foundation of China (NSFC) under Contract No.
11675200}\\
        (On behalf of the BESIII Collaboration)\\
        Institute of High Energy Physics, CAS, Beijing, China, 100049\\
        E-mail: \email{chenjc@ihep.ac.cn}}

\abstract{
We present recent BESIII results about the
study of mesons which contain at least one charm quark.
We determined the $D_{(s)}^{+}$ decay constants, the form factors of $D$
semi-leptonic decays, the CKM matrix elements $|V_{cs(d)}|$.
}

\FullConference{ICHEP2018, XXXIX International Conference on 
High Energy Physics \\
                4-11 July 2018\\
                Seoul, Korea}

\begin{document}

\section{Introduction}
The study of mesons which contain at least one charm quark is 
referred to as open charm physics. It offers the possibility to study up-type
quark transitions. Since the $c$ quark cannot be treated in any mass limit,   
theoretical predictions are difficult and experimental input is crucial.
The at-threshold decay topology offers special opportunities to study open 
charm decays. 

The BESIII is a magnetic spectrometer working at a double-ring
$e^+e^-$ collider operating at
center-of-mass energy between 2.0 GeV and 4.6 GeV,
located at the Institute of High Energy Physics (IHEP) in Beijing.
The maximum luminosity of the BEPCII at 3.773 GeV is
$1\times 10^{33}\rm cm^{-2}\rm s^{-1}$~\cite{BESIII}. 
The samples of interest for the
analysis described in the following were taken at the $D\bar{D}$ threshold
(3.773 GeV) and $D^*_sD_s$ threshold (4.178 GeV) with integrated
luminosities of 2.92 fb$^{-1}$ and 3.19 fb$^{-1}$, respectively. 
Throughout in the following, charge conjugate is implied.

\section{\boldmath Pure leptonic $D_{(s)}$ decays}
The pure leptonic decay of charged $D_{(s)}^+$ mesons proceeds via 
the annihilation of the $c$ quark and the anti-$d$ (anti-$s$) quark to
a virtual $W^{\pm}$ boson and its decay to $\ell^+\nu_\ell$. 
The decay rate can be parameterized as:
\begin{equation}
\Gamma(D_{(s)}^+\to\ell^+\nu_\ell)=\frac {G_{F}^2}{8\pi}f^2_{D_{(s)}}m^2_\ell m_{D_{(s)}}
(1-\frac {m^2_\ell}{m^2_{D_{(s)}}})^2|V_{cd(s)}|^2,
\label{DecayRate}
\end{equation}
where $G_{F}$ is the Fermi coupling constant, 
$m_\ell$ and $m_{D_{(s)}}$ are
the masses of the lepton and the $D_{(s)}$ meson in the final state, respectively.

\subsection{$D^+\to\mu^{+}\nu_{\mu}$}
\begin{figure}
\begin{center}
\parbox{6.0cm}{\includegraphics[width=5.5cm,height=5.5cm ] {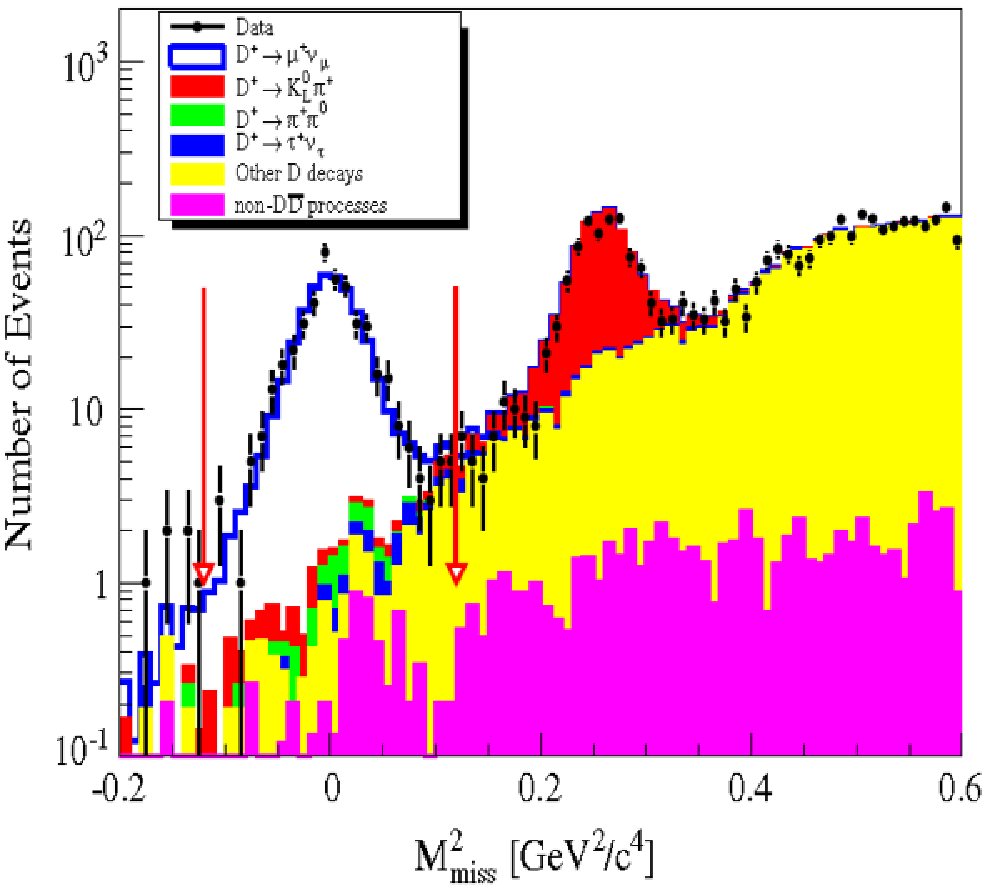}
\caption{\label{fig:MM_miss_Duv}
  The $M_{\rm miss}^2$ distributions of the
  $D^{+}\to \mu^{+}\nu_{\mu}$ candidates,
  where two arrows denote the signal region.
  }}
\hspace{1.0cm}
\parbox{6.0cm}{\includegraphics[width=6.0cm,height=6.0cm]{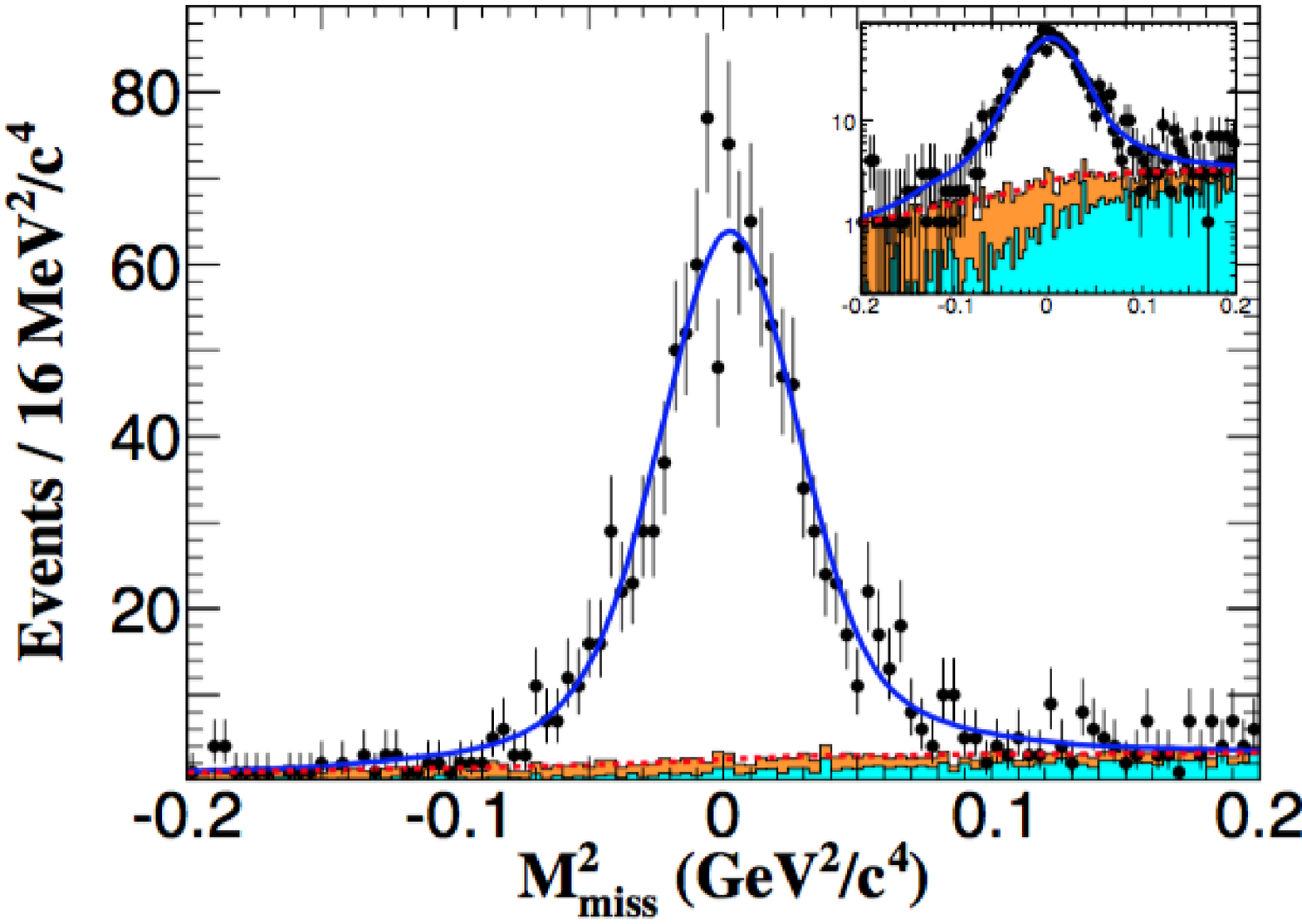}
 \caption{ \label{fig:MM_miss_Dslep}
  Fit of the accepted the $D_s^{+}\to \mu^{+}\nu_{\mu}$ candidates events 
(BESIII preliminary result).
  }}
\end{center}
\end{figure}

By analyzing the data sample accumulated at 3.773 GeV, we first selected
the $D^-$ meson sample from its hadronic decay modes, which is called
singly tagged (ST) $D^-$. The $D^{+}\to\mu^{+}\nu_{\mu}$ decays are selected
in the 
recoil side of the ST $D^{-}$. To select the leptonic $D^{+}$ decay
with a missing neutrino, we calculated $U_{\rm miss}=E_{\rm miss}-p_{\rm
miss}$ or $M_{\rm miss}^2=E_{\rm miss}^2-p_{\rm miss}^2$, 
where $E_{\rm miss}$ and $p_{\rm miss}$ are the missing energy and missing 
momentum of the event.
Figure~\ref{fig:MM_miss_Duv} shows the distribution of $M_{\rm miss}^2$.  
After subtracting the backgrounds, $409.0\pm21.2$ signal
events are retained and the branching fraction (BF) is measured to be
$BF(D^{+}\to 
\mu^{+}\nu_{\mu})=(3.71\pm0.19_{stat}\pm0.06_{syst})\times 10^{-4}$,
which is the most precise measurement to date.
Combining this BF measurement and the Particle Data Group (PDG)~\cite{PDG}
values of $D^+$ lifetime, $m_{D^+}$, $m_{\mu}$ and magnitude of $|V_{cd}|$ 
determined from the global Standard Model (SM) fit, the
decay constant is determined to be $f_D = (203.2\pm5.3_{stat}\pm1.8_{syst})$
MeV. Alternatively, the magnitude
of CKM matrix element |V$_{cd}$| is extracted to be 
|V$_{cd}$| = $0.2210\pm0.0058_{stat}\pm0.0047_{syst}$~\cite{Dmuv}.

\subsection{$D_s^+\to\mu^+\nu_\mu$}
By analyzing the data taken at 4.178 GeV, we also studied the leptonic $D^+_s$ decays.
From 14 $D^-_s$ hadronic decay modes, preliminarily, $388660 \pm 2592$ ST 
$D_s^-$ mesons were accumulated. In the system recoiling against the ST 
$D_s^-$, 
the signal events of $D_s^+\to\mu^+\nu_\mu$ decay were selected. 
The distribution of $M_{\rm miss}^2$ is shown in Fig.~\ref{fig:MM_miss_Dslep}. 
We obtained $1135.9\pm33.1$ $D_s^+\to\mu^+\nu_\mu$, and
the preliminary result of absolute BF is determined to be
$BF(D_s^{+}\to \mu^{+}\nu_{\mu}) = 
(0.550\pm0.016_{stat}\pm0.015_{syst})$\%,
the decay constant $f_{D_s} = 245.9\pm3.6_{stat}\pm3.5_{syst}$ MeV and 
$|V_{cs}|=0.987\pm0.014_{stat}\pm0.015_{syst}$. The additional systematic uncertainties
according to the input parameters are negligible for |V$_{cs}$| and 0.3 
for $f_{D_s}$.

\section{\boldmath Semi-leptonic $D_{(s)}$ decays}
\subsection{$D^{0(+)} \to P e^+\nu_e$ ($P = K^-,\pi^-, K^0, \pi^0$)}
In the SM, neglecting the lepton mass, the differential decay rate for 
$D^{0(+)} \to P e^+\nu_e$ ($P = K^-,\pi^-, K^0$ or $\pi^0$) is given by
\begin{equation}
\frac{d\Gamma}{dq^2}= X \frac {G_{F}^2}{24\pi^3}|V_{\rm cd(s)}|^2 p^3 |f^P_+(q^2)|^2,
\label{SLDecayRate}
\end{equation}
where $X$ is a multiplicative factor due to isospin,
which equals to 1 for modes $D^0 \to K^- e^+\nu_e$, $D^0 \to \pi^- e^+\nu_e$,
$D^+ \to K^0 e^+\nu_e$, and 1/2 for mode $D^+ \to\pi^0 e^+ \nu_e$, 
$p$ is the momentum of the pseudo-scalar meson $P$ in the rest frame of the $D$ meson, 
$q^2$ is the squared four momentum transfer, i.e., the invariant mass of the electron and 
neutrino system, 
$f^P_+(q^2)$ is the form factor which describes the strong interaction between the final state quarks and
is usually parameterized in data analysis. 

Based on the data taken at 3.773 GeV, BESIII studies the dynamics
of the $D^0 \to K^- e^+\nu_e$ and $D^0\to\pi^- e^+\nu_e$ decays. 
The BFs are measured to be
$BF(D^0 \to K^- e^+\nu_e)=(3.505\pm0.014_{stat}\pm 0.033_{syst})$\% 
and $BF(D^0 \to \pi^- e^+\nu_e)=
(0.295\pm0.004_{stat}\pm0.003_{syst})$\%~\cite{D0kpiev}. Similarly, 
the BFs are measured to be 
$BF(D^+ \to \bar K^0 e^+\nu_e) = (8.604\pm0.056_{stat} \pm0.151_{syst})$\% 
and $BF(D^+ \to \pi^0 e^+\nu_e) = 
(0.363\pm0.008_{stat}\pm0.005_{syst})$\%~\cite{Dk0pi0ev}.

\begin{figure}
\begin{center}
\includegraphics[width=7.0cm]{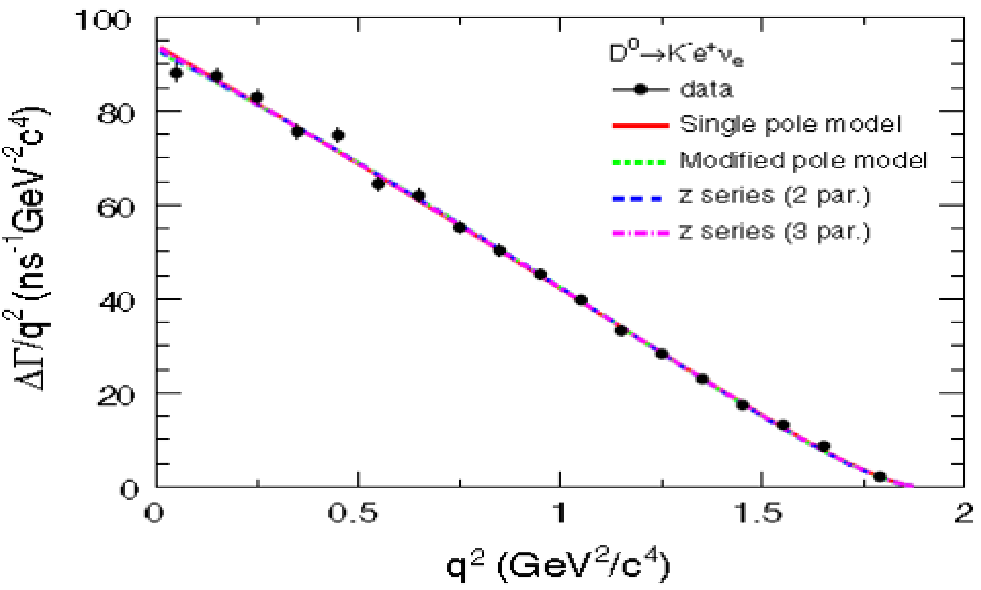}
\includegraphics[width=7.0cm]{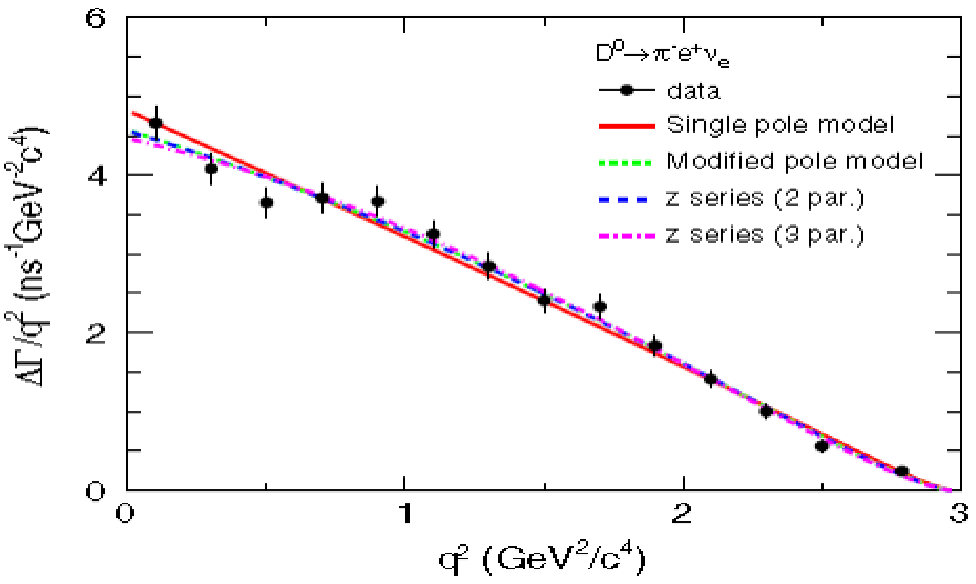}
 \caption{ \label{fig:q2_D0kpiev}
  Fits to partial decay widths of $D^0 \to K^- e^+\nu_e$
(left) and $D^0 \to \pi^- e^+\nu_e$ (right).
  }
\end{center}
\end{figure}

\begin{figure}
\begin{center}
\includegraphics[width=14.0cm]{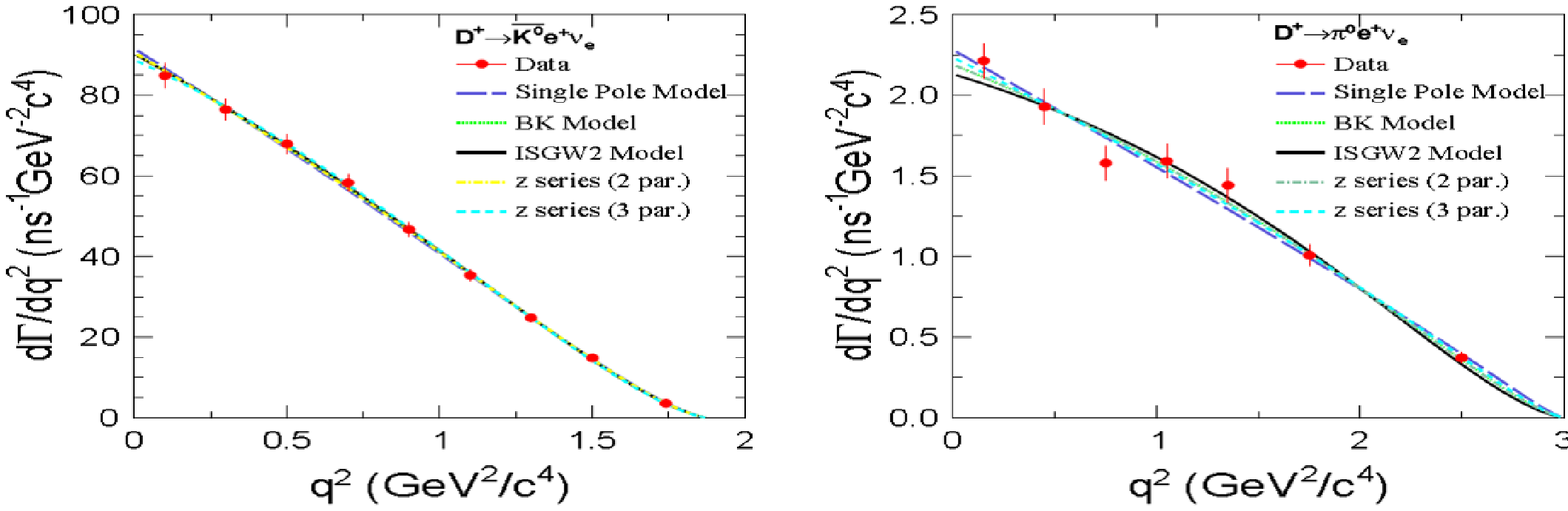}
 \caption{ \label{fig:q2_Dkpiev}
  Fits to partial decay widths of $D^+ \to \bar K^0 e^+\nu_e$
(left) and $D^+ \to \pi^0 e^+\nu_e$ (right).
  }
\end{center}
\end{figure}

We also studied the differential decay rates of these two processes. 
Figures~\ref{fig:q2_D0kpiev} and ~\ref{fig:q2_Dkpiev} show the fit results. 
We extract the product $f_+(0)|V_{cs(d)}|$ and
other form factor parameters.
Using the values for $f_+^{K(\pi)}(0)|V_{cs(d)}|$
from the two-parameter $z$-series expansion fits and with
$f^K_+(0) = 0.747 \pm 0.011 \pm 0.015$~\cite{fk0} and 
$f^{\pi}_+ (0) = 0.666 \pm0.020 \pm 0.021$~\cite{fpi0}
calculated in LQCD, $|V_{cs}|$ is obtained
to be $0.9601 \pm 0.0033 \pm 0.0047 \pm 0.0239$ ($|V_{cd}| =
0.2155\pm0.0027\pm0.0014\pm0.0094$), where the first uncertainties
are statistical, the second ones systematic, and
the third ones are due to the theoretical uncertainties in
the form factor calculations.
The BESIII
results are in good agreement with the previous measurements,
and with the best precision to date.


\subsection{$D^{0(+)} \to P \mu^+\nu_\mu$ ($P = K^-,\pi^-, K^0, \pi^0$)}
Based on the data sample collected at $\sqrt{s}=3.773$ GeV, BESIII although
studied $D^{0(+)} \to P \mu^+\nu_\mu$ ($P = K^-,\pi^-, K^0, \pi^0$).
The BFs and the lepton universality ratios 
$R_{LU}=BF(D^{0(+)} \to P \mu^+\nu_\mu)/BF(D^{0(+)} \to P e^+\nu_e)$
are measured to be 
$BF(D^0 \to K^- \mu^+\nu_{\mu}) = (3.429\pm0.019_{stat} \pm0.035_{syst})$\% 
($R_{LU}=0.978\pm0.007_{stat}\pm0.012_{syst}$)~\cite{D0kmuv},   
$BF(D^+ \to \bar K^0 \mu^+\nu_{\mu}) = (8.72\pm0.07_{stat} \pm0.18_{syst})$\% 
($R_{LU}=0.988\pm0.033$)~\cite{Dpk0muv},  
$BF(D^0 \to \pi^- \mu^+\nu_{\mu}) = (0.267\pm0.007_{stat} \pm0.007_{syst})$\% 
($R_{LU}=0.905\pm0.027_{stat}\pm0.023_{syst}$) and  
$BF(D^+ \to \pi^0 \mu^+\nu_{\mu}) = (0.342\pm0.011_{stat} \pm0.010_{syst})$\% 
($R_{LU}=0.942\pm0.037_{stat}\pm0.027_{syst}$)~\cite{D2pimuv}.  

\subsection{$D_s^{+} \to K^{(*)0}e^+\nu_e$}
Using the data sample collected at $\sqrt{s}=4.178$ GeV, BESIII 
measured $D_s^{+} \to K^{0}e^+\nu_e$ and $D_s^{+} \to K^{*0}e^+\nu_e$. 
The BFs are obtained to be
$BF(D_s^+ \to K^0 e^+\nu_{e}) = (3.25\pm0.38_{stat} \pm0.16_{syst})$\% and  
$BF(D_s^+ \to K^{*0} e^+\nu_{e}) = (2.37\pm0.26_{stat} \pm0.20_{syst})$\%
~\cite{Dskstmuv}.
The first measurements of the hadronic form-factor parameters are obtained.
The result for $D_s^{+} \to K^{0}e^+\nu_e$ is
$f_+(0)=0.720\pm0.084(stat)\pm0.013(syst)$, and for 
$D_s^{+} \to K^{*0}e^+\nu_e$ decay, the form-factor ratios
are $r_V=V(0)/A_1(0)=1.67\pm0.34(stat)\pm0.016(syst)$ and 
$r_2=A_2(0)/A_1(0)=0.77\pm0.28(stat)\pm0.07(syst)$.

\subsection{\boldmath $D^{+}_s \to \eta(\eta') e^+\nu_e$}
By analyzing the data taken at 4.178 GeV, 
BESIII measured the absolute BFs for semi-leptonic 
$D^{+}_s \to \eta(\eta') e^+\nu_e$ decays.
The preliminary results are
$BF(D^{+}_s \to \eta e^+\nu_e) = (2.32\pm0.06_{stat} 
\pm0.06_{syst})$\%
and $BF(D^{+}_s \to \eta' e^+\nu_e) = (0.82\pm0.07_{stat} 
\pm0.03_{syst})$\%, and
combining the BFs of $BF_{D^{+} \to \eta(\eta') e^+\nu_e}$, the
$\eta-\eta'$ mixing angle is determined to be 
$\phi_P=(40.2\pm1.4_{stat}\pm0.5_{syst})^\circ$. 
From the first measurements of the dynamics of $D_s^+\to \eta(')e^+\nu_e$ 
decays, the products of the hadronic form factors $f_+^{\eta(')}$(0) and 
the CKM matrix element |V$_{cs}$| are extracted with different form factor
parameterizations. For the two parameter series expansion, the preliminary
results are $f_{\eta}$(0)|V$_{cs}$| = 
$0.446\pm0.005_{stat}\pm0.004_{syst}$
and $f_+^{\eta'}$(0)|V$_{cs}$|=$0.477\pm0.049_{stat}\pm0.011_{syst}$. 
Taking
|V$_{cs}$| from the CKMfitter as input, we determine preliminary
$f_+^{\eta}(0) = 0.458\pm0.005_{stat}\pm0.004_{syst}$ and 
$f_+^{\eta'}(0) = 0.490\pm0.050_{stat}\pm0.011_{syst}$. Alternatively, 
using the $f_+^{\eta(')}$(0) calculated by light-cone sum rules
leads to |V$_{cs}$|=$1.032\pm0.012_{stat}\pm0.009_{syst}\pm0.079_{theo}$ 
and $0.917\pm0.094_{stat}\pm0.021_{syst}\pm0.155_{theo}$, respectively.

\section{ Summary}
We present a selection of recent BESIII charm results
based on the data sets collected by BESIII detector near 
the $D\bar{D}$ threshold (3.773 GeV), $D^*_sD_s$ threshold (4.178 GeV)
with integrated luminosities of 2.93 fb$^{-1}$ and 3.19 fb$^{-1}$, 
respectively.
From the leptonic $D_{(s)}$ pure leptonic and semi-leptonic decays we 
determined 
the most precise values for the decay constant $f_{D^+_{(s)}}$, the 
hadronic form factors $f^{ K (\pi)}_+(0)$, and the form factor shape 
$f^{ K (\pi)}_+(q^2)$
which provide important test to LQCD calculations, and CKM matrix unitary.

\end{document}